\documentclass[eprint, aps, prl, twocolumn, superscriptaddress, showpacs, floatfix]{revtex4-1}
\usepackage{amssymb}
\usepackage{mathrsfs} % cal F fidelity
\usepackage{amsmath}
\usepackage{float}
\usepackage{bbm}
\usepackage{graphicx}
\usepackage{epsfig}
\usepackage{verbatim}
\usepackage{bm}
\usepackage[usenames]{color}

\usepackage{multirow}
\usepackage{extarrows}
\usepackage{array}

\usepackage{hyperref}
\begin{document}

\title{Randomized benchmarking of barrier versus tilt control of a singlet-triplet qubit}

\author{Chengxian Zhang}
\affiliation{Department of Physics and Materials Science, City University of Hong Kong, Tat Chee Avenue, Kowloon, Hong Kong SAR, China}
\affiliation{City University of Hong Kong Shenzhen Research Institute, Shenzhen, Guangdong 518057, China}
\author{Robert E. Throckmorton}
\affiliation{Condensed Matter Theory Center and Joint Quantum Institute, Department of Physics, University of Maryland, College Park, Maryland 20742, USA}
\author{Xu-Chen Yang}
\affiliation{Department of Physics and Materials Science, City University of Hong Kong, Tat Chee Avenue, Kowloon, Hong Kong SAR, China}
\affiliation{City University of Hong Kong Shenzhen Research Institute, Shenzhen, Guangdong 518057, China}
\author{Xin Wang}
\email{x.wang@cityu.edu.hk}
\affiliation{Department of Physics and Materials Science, City University of Hong Kong, Tat Chee Avenue, Kowloon, Hong Kong SAR, China}
\affiliation{City University of Hong Kong Shenzhen Research Institute, Shenzhen, Guangdong 518057, China}
\author{Edwin Barnes}
\affiliation{Department of Physics, Virginia Tech, Blacksburg, Virginia 24061, USA}
\author{S.~Das Sarma}
\affiliation{Condensed Matter Theory Center and Joint Quantum Institute, Department of Physics, University of Maryland, College Park, Maryland 20742, USA}
\date{\today}

\begin{abstract}
Decoherence due to charge noise is one of the central challenges in using spin qubits in semiconductor quantum dots as a platform for quantum information processing. Recently, it has been experimentally demonstrated in both Si and GaAs singlet-triplet qubits that the effects of charge noise can be suppressed if qubit operations are implemented using symmetric barrier control instead of the standard tilt control. Here, we investigate the key issue of whether the benefits of barrier control persist over the entire set of single-qubit gates by performing randomized benchmarking simulations. We find the surprising result that the improvement afforded by barrier control depends sensitively on the amount of spin noise: for the minimal nuclear spin noise levels present in Si, the coherence time improves by more than two orders of magnitude whereas in GaAs, by contrast the coherence time is essentially the same for barrier and tilt control. However, we establish that barrier control becomes beneficial if qubit operations are performed using a new family of composite pulses that reduce gate times by up to 90\%. With these optimized pulses, barrier control is the best way to achieve high-fidelity quantum gates in singlet-triplet qubits.
\end{abstract}

\maketitle

Efficient and robust control of quantum systems is key to quantum information processing. Using spins in semiconductor quantum dots is a promising approach due to their fast operation times, long coherence times, and prospect for scalability \cite{Hanson.07,Zwanenburg.13}. Much progress has been made in designing and demonstrating high-fidelity control over single- and multi-qubit devices based on various qubit types, including single-spin qubits \cite{Loss.98,Nowack.11,Pla.12,Pla.13,Veldhorst.14,Braakman.13,Otsuka.16,ItoArxiv2016},
double-dot singlet-triplet qubits \cite{Levy.02,Petta.05,Foletti.09,vanWeperen.11,Maune.12,Shulman.12,Dial.13,Shulman.14,Reed.16,Martins.16},
triple-dot exchange-only qubits \cite{DiVincenzo.00,Medford.13a,Medford.13b,Eng.15,Shim.16}, and
``hybrid'' qubits \cite{Shi.12,Kim.14,Kim.15}.

Singlet-triplet qubits \cite{Levy.02,Petta.05} are particularly promising due to their relatively simple all-electrical control and long coherence times \cite{Bluhm.11,Malinowski.16}. Achieving high-fidelity gate operations, however, has been challenging due primarily to charge noise. Qubit operations are implemented by tuning the Heisenberg exchange interaction between the two electron spins, and charge fluctuations in the vicinity of the dots introduce noise into this interaction, causing decoherence \cite{Hu.06,Dial.13,Barnes.16,Throckmorton.16}. Until recently, the standard method for tuning this interaction was to tilt the electrostatic potential defining the two quantum dots by raising the chemical potential in one dot relative to the other \cite{Petta.05}, a method that inherently treats the two dots asymmetrically. In recent groundbreaking experimental works \cite{Bertrand.15,Reed.16,Martins.16}, it was shown that if symmetry is maintained between the dots and the exchange interaction is tuned instead by adjusting the electrostatic barrier separating them, then the sensitivity of the qubit to charge noise can be reduced substantially. This is an important result not only for singlet-triplet qubits, but for quantum dot spin qubits in general given the central role of the exchange interaction and the prevalence of charge noise in all such qubits.

Despite its importance, the true benefit of symmetric barrier control is yet to be determined. This is because the advantages of this technique have only been examined for a very limited set of gate operations, and how it will perform in quantum algorithms involving many types of gates remains completely unknown. This is a subtle issue because the effect of charge noise depends on (and increases with) the strength of the exchange interaction, which in turn depends on the specific gate being implemented. Additional noise sources, such as nuclear spin noise \cite{Medford.12}, can thus become more important than charge noise for certain gates. Therefore, the performance of barrier control versus tilt control depends sensitively on the control sequences used to implement gates, and this dependence must be understood in order to take full advantage of barrier control.

In this paper, we perform randomized benchmarking \cite{Magesan.12} on singlet-triplet qubits to quantitatively determine the improvement afforded by symmetric barrier control. We extract the effective single-qubit gate fidelity by averaging over all single-qubit Clifford gates for both barrier and tilt control, using experimentally measured levels of charge noise in each case. We perform this comparison for varying levels of nuclear spin noise ranging from zero (purified Si) to typical values for GaAs. Using standard gate control sequences, we find that in the absence of nuclear spin noise (i.e., in Si), barrier control improves the coherence time by more than two orders of magnitude. On the other hand, for typical levels of nuclear spin noise in GaAs, barrier control gives essentially no improvement because nuclear spin noise dominates for most gates in the set. To overcome this problem, we present a new set of composite pulses that implement gates up to ten times faster. These pulses reduce the sensitivity to nuclear spin noise, allowing the benefits of barrier control to become visible in benchmarking simulations even for high levels of nuclear spin noise.

In the case of singlet-triplet qubits, universal operations require a magnetic field gradient across the two dots in addition to the tunable exchange interaction. This gradient can be produced either by nuclear spins \cite{Foletti.09} or a micromagnet \cite{Wu.14}, and in both cases it remains fixed during gate operations. The qubit Hamiltonian can be written in the form $H(t)=J(t)\sigma_z+h\sigma_x$, with $J(t)$ denoting the exchange interaction and $h$ the field gradient \cite{Petta.05}. Charge noise and nuclear spin noise can be included as stochastic fluctuations $\delta J$, $\delta h$ in $H(t)$. Nuclear spin noise tends to be much more significant in GaAs compared with Si, however, its influence on qubit coherence can be controlled in both materials: in Si through isotopic purification, and in GaAs through nuclear spin programming \cite{Foletti.09,Bluhm.10} and Bayesian estimation \cite{Shulman.14}. Both types of noise are known to have power-law frequency spectra \cite{Medford.12,Dial.13}.  

Randomized benchmarking \cite{Knill.08,Magesan.12} is a powerful technique to extract the average gate fidelity using the Clifford group, a subset of all possible single-qubit gates. The randomized benchmarking procedure is implemented by averaging the
fidelity \cite{Bowdrey.02} over random sequences of single-qubit Clifford gates,
and over different noise realizations. We consider both the full frequency-dependent noise model and the simpler `quasistatic' model in our simulations. The latter assumes noise fluctuations are constant for each run of the experiment (varying from one run to the next) and is valid for sufficiently short qubit operations. In order to perform the averaging for either noise model, we must first determine how to implement arbitrary single-qubit gates in singlet-triplet qubits.

\begin{figure}
	\includegraphics[width=\columnwidth]{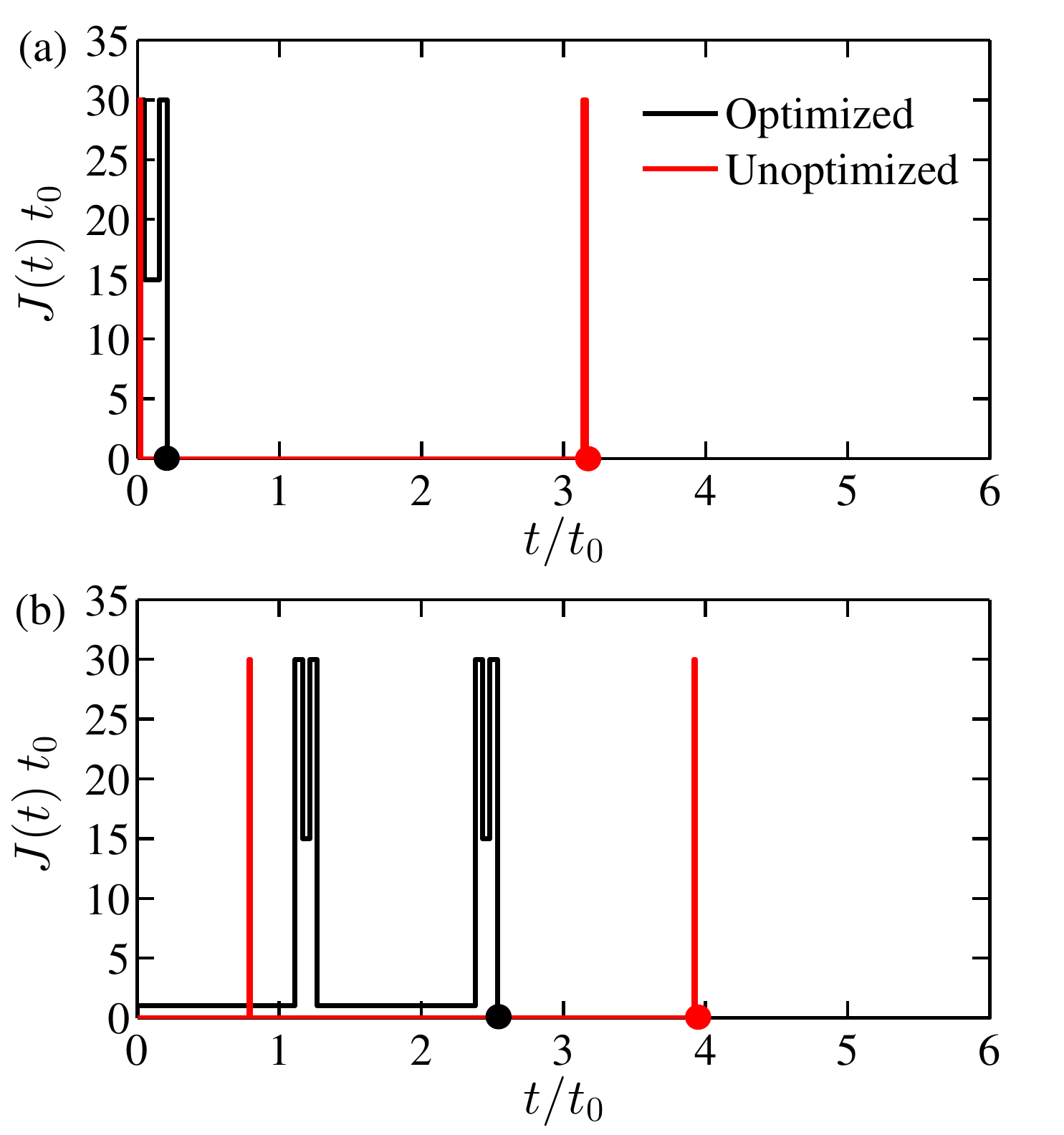}
	\caption{Pulse profiles of the unoptimized (red/gray lines) and optimized (black lines) pulse sequences for (a) $R(\hat{z},\pi)$ and (b) $R(\hat{x}+\hat{y}+\hat{z},2\pi/3)$.  Here, $t_0$ is a time scale that is taken to be $1/h$ throughout this paper.}
	\label{fig:pulseshape}
\end{figure}

The simplest operations generated by $H(t)$ are rotations of the Bloch vector around axes lying in the $xz$ plane, $R(h\hat{x}+J\hat{z},\phi)$, where the first entry indicates the rotation axis determined by the ratio $J/h$, and the second entry is the rotation angle, which is determined by $\sqrt{J^2+h^2}$ and the operation time. These gates can be performed in a single shot, meaning that $J(t)$ is fixed to one value throughout the operation. All other types of single-qubit gates must be implemented using composite pulses, in which $J(t)$ assumes a few different values over the course of the gate. These can be designed by invoking the well known fact that an arbitrary rotation $R(\hat{r},\phi)$ can be decomposed into an $x$ rotation sandwiched between two $z$ rotations (``the $z$-$x$-$z$ sequence'') \cite{NielsenChuang.00,Wang.14a}:
\begin{equation}
R(\hat{r},\phi)=R(\hat{z},\phi_1)R(\hat{x},\phi_2)R(\hat{z},\phi_3),\label{eq:zxzseq}
\end{equation}
where $\phi_{1,2,3}$ are auxiliary angles depending on the desired rotation. Since in practice the magnetic field gradient cannot be turned on and off during a given gate operation, $z$ rotations must be further broken down as (``the Hadamard-$x$-Hadamard sequence'') \cite{NielsenChuang.00}
\begin{equation}
R(\hat{z},\phi)=-R(\hat{x}+\hat{z},\pi)R(\hat{x},\phi)R(\hat{x}+\hat{z},\pi).\label{eq:HxHseq}
\end{equation}
In practice it is more convenient to use a generalized version of Eq.~\eqref{eq:HxHseq} (``the Ramon sequence'') \cite{Ramon.11}:
\begin{equation}
R(\hat{z},\phi)=R(\hat{x}+\cot\theta\hat{z},\chi)R(\hat{x},\alpha)R(\hat{x}+\cot\theta\hat{z},\chi), \label{eq:Ramon}
\end{equation}
where $\chi$ and $\alpha$ depend on $\phi$ and $\theta$. Any single-qubit gate can be implemented in either a single shot or using one of these composite pulse sequences. Two examples have been shown as red lines (``unoptimized'') in Fig.~\ref{fig:pulseshape}. The black lines are examples of our new (``optimized'') sequences that speed up gates by using an alternative to the standard Ramon sequence, as we explain below.

\begin{figure}
	\includegraphics[width=\columnwidth]{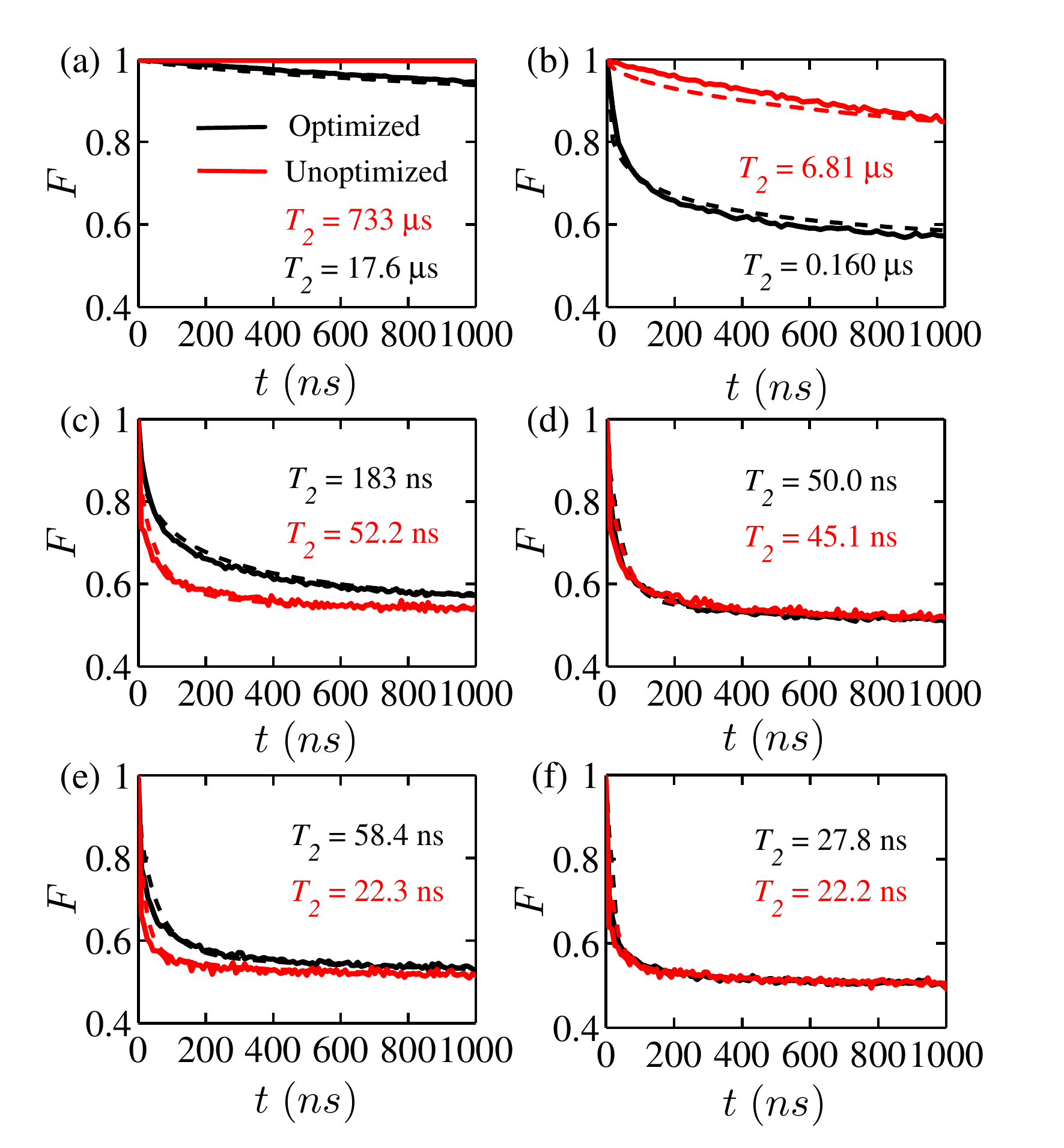}
	\caption{Results of randomized benchmarking with Gaussian noise (i.e., the quasistatic limit) for unoptimized (red) and optimized (black) pulses. The dashed lines are exponential fits as detailed in the main text.
	The magnetic field gradient and nuclear spin noise values are $h=23$ MHz and $\delta h=0$ for (a) and (b), $h=40$ MHz and $\sigma_h=11.5$ MHz for (c) and (d), and $h=40$ MHz and $\sigma_h=23$ MHz for (e) and (f).  For the left column [(a), (c), (e)], we use barrier control, and thus we take $\sigma_J=0.00426 J$, while we use tilt control in the right column [(b), (d), (f)] and thus we use $\sigma_J=0.0563 J$. We indicate the fitted $T_2$ values on this figure in the same color as the lines that they correspond to, and we provide the fitting parameters
	 in the Supplemental Material \cite{suppl}.}
	\label{fig:static}
\end{figure}

For our randomized benchmarking simulations, we first consider the Gaussian quasistatic model employed in Refs.~\cite{Martins.16,Barnes.16,Throckmorton.16} to characterize both barrier and tilt control. In this model, $\delta h$ is static for each run and drawn from a Gaussian distribution with zero mean and variance $\sigma_h^2$: ${\cal N}(0,\sigma_h^2)$, while $\delta J$ is proportional to $J$, with the proportionality constant drawn from ${\cal N}(0,\sigma_J^2/J^2)$ for each piece of the control pulse. The relation between $\delta J$ and $J$ arises directly from the experimentally measured dependence of $J$ on applied voltage \cite{Shulman.12,Dial.13,Martins.16}. In our simulation of charge noise, we use the experimentally fitted value $\sigma_J/J=0.00426$ for barrier control and 0.0563 for tilt control \cite{Martins.16}. The latter value reflects typical experimental exchange field fluctuations at the few percent level \cite{Dial.13}. Here and throughout this work, we averaged results from 2000 random Clifford gate sequences and noise realizations to ensure convergence. For most results, fitting to $\left\{1+\exp\left[-(t/T_2)^\gamma\right]\right\}/2$ suffices, where $T_2$ is an effective coherence time quantifying the performance of the control scheme. However in some cases a better fit is achieved using a summation of two exponentials $\left\{2+\exp\left[-(t/T_2)^{\gamma_1}\right]+\exp\left[-(t/T_2)^{\gamma_2}\right]\right\}/4$.

The red curves in Fig.~\ref{fig:static} are our results for the average fidelity as a function of qubit evolution time using composite pulses built from the (unoptimized) Ramon sequence. We have chosen $J_{\rm max}/h=\cot\theta=30$ throughout this paper, and we have also verified that our conclusions are identical for other experimentally relevant values $J_{\rm max}\gtrsim 10h$. Panels in the left column correspond to symmetric barrier control, while those on the right correspond to tilt control. Figs.~\ref{fig:static}(a),(b) show the case without nuclear spin noise, $\delta h{=}0$, where it is evident that $T_2$ increases by more than two orders of magnitude when barrier control is used instead of tilt. Thus, in Si systems where nuclear spin noise is minimal, barrier control provides a very large improvement, as is consistent with Ref.~\cite{Reed.16}. In contrast, Figs.~\ref{fig:static}(c),(d) and (e),(f) show results for moderate ($\sigma_h{=}11.5$ MHz) and high ($\sigma_h{=}23$ MHz) levels of nuclear spin noise; these values are typical for GaAs, with the latter taken from Ref.~\cite{Martins.16}. In these cases, barrier control shows little improvement over tilt control. We attribute this to the fact that the Ramon sequence, Eq.~\eqref{eq:Ramon}, includes a segment with $J{=}0$. This simultaneously reduces the amount of charge noise (since $\delta J\sim J$) and slows down gates, increasing exposure to nuclear spin noise. Thus when standard pulse sequences are used, barrier control is only effective if nuclear spin noise is strongly suppressed (i.e., Si but not GaAs).

To take greater advantage of barrier control when nuclear spin noise is significant, we need to avoid setting $J=0$ as much as possible. We can achieve this by using new (optimized) composite pulses based on the identity
\begin{equation}
R(\hat{z},\phi){=}-R(\hat{x}+\cot\theta\hat{z},\pi)R(\hat{x}+\cot2\theta\hat{z},\phi)R(\hat{x}+\cot\theta\hat{z},\pi), \label{eq:Robert}
\end{equation}
where a pure $x$ rotation is no longer needed. Figure~\ref{fig:pulseshape}(a) compares a realization of $R(\hat{z},\pi)$ using the optimized sequence \eqref{eq:Robert} (black line)  and the unoptimized one \eqref{eq:Ramon} (red/gray line);  we see that the gate time has been substantially reduced by a factor of ${\gtrsim} 10$. 
Sandwiching Eq.~\eqref{eq:Robert} between two Hadamard gates results in an $x$ rotation \cite{xrot}. Arbitrary rotations can therefore be done using the composite $z$ and $x$ rotations discussed above according to Eq.~\eqref{eq:zxzseq}. Fig.~\ref{fig:pulseshape}(b) shows an example comparing the optimized and unoptimized sequences achieving $R(\hat{x}+\hat{y}+\hat{z},2\pi/3)$, and one can clearly see that the optimized sequence is ${\sim}40\%$ shorter than the unoptimized one. We have constructed 24 single-qubit Clifford gates similarly and have found that the reduction of gate time is between 40\% and 60\% for all gates except direct $z$ and $x$ rotations \cite{suppl}. 

The black curves in Fig.~\ref{fig:static} show our results for randomized benchmarking using our new optimized pulses. Now we see that for all levels of nuclear spin noise, barrier control is superior to tilt control. In the absence of nuclear spin noise (panels (a),(b)), the unoptimized pulses achieve longer coherence times than the optimized ones since they are less sensitive to charge noise. However, in the presence of nuclear spin noise, the optimized pulses show longer coherence times. This improvement is due directly to the fact that our optimized pulses are considerably faster. Even in the absence of nuclear spin noise, the optimized pulses may be preferable due to their shorter durations \cite{Reed.16}.

To test our new optimized pulse sequences, we carry out randomized benchmarking with the frequency dependence of the noise {\it included} \cite{Yang.16}. For both charge and spin noise, this dependence has been measured to be of $1/f$ type \cite{Medford.12,Dial.13}, i.e. the nuclear spin (indicated by $h$) and charge noises ($J$) exhibit spectra $S_{h,J}(\omega)=A_{h,J}/(\omega t_0)^{\alpha_{h,J}}$, where $A$ is the amplitude, the exponent $\alpha_{h,J}$ signifies the self-correlation of the nuclear ($h$) and charge ($J$) noise respectively, and $t_0$ is the time unit, taken as $t_0=1/h$. When nuclear spin noise is absent, one simply sets $A_h=0$ and $\delta h=0$. However, in order to map to the quasistatic model with $\sigma_h=23$ MHz used in Ref.~\cite{Martins.16}, we mandate that the integrated $1/f$ noise, $\int_{\omega_{\rm ir}}^{\omega_{\rm uv}}d\omega A_h/(\omega t_0)^{\alpha_h}$, equals that of the quasistatic Gaussian noise $\pi\sigma_h^2$ \cite{Barnes.16}. To facilitate the discussion we will simply refer to this case as ``$\delta h\neq0$'' in the remainder of this paper. The low and high frequency cutoffs are taken as $\omega_{\rm ir}=10$ kHz and $\omega_{\rm uv}=100$ kHz \cite{Barnes.16}. Similarly the charge noise amplitude, $A_J$, is determined by solving
\begin{equation}
\int_{\omega_{\rm ir}}^{\omega_{\rm uv}} \frac{A_J}{(\omega t_0)^{\alpha_J}}d\omega=\pi\left(\frac{\sigma_J}{Jt_0}\right)^2
\end{equation}
for the two cases $\sigma_J=0.00426J$ and $\sigma_J=0.0563J$, corresponding to barrier and tilt control, respectively. The cutoffs are taken as $\omega_{\rm ir}=50$ kHz and $\omega_{\rm uv}=1$ MHz \cite{Barnes.16}.

\begin{figure}
	\includegraphics[width=\columnwidth]{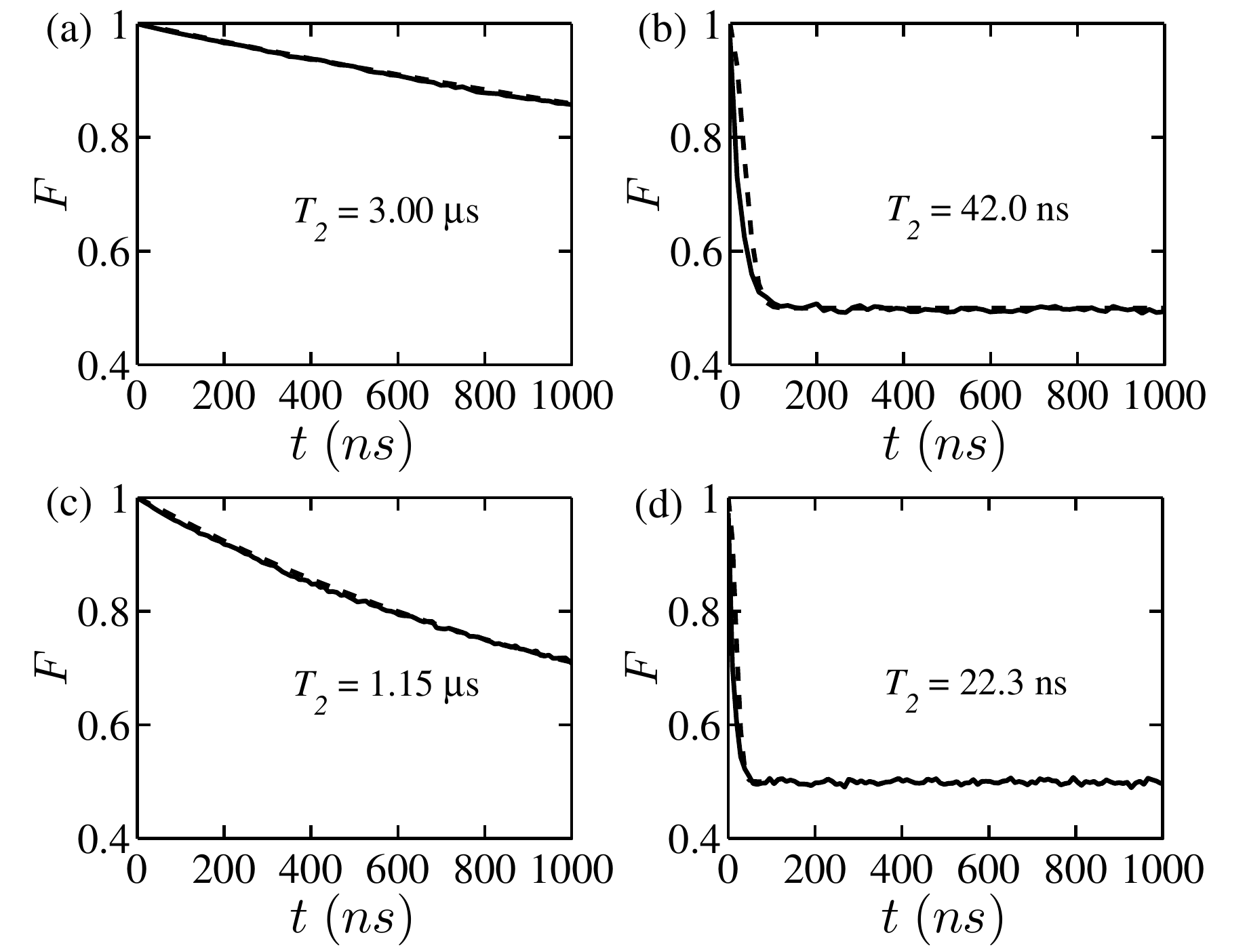}
	\caption{
	Randomized benchmarking results of the optimized pulse sequences under $1/f^{2.6}$ nuclear noise and $1/f^{0.7}$ charge noise (i.e. $\alpha_h = 2.6$ and $\alpha_J=0.7$).  The solid curves are our numerical results and the dashed curves are fits.  (a) and (b) correspond to $h = 23$ MHz and $\delta h=0$, while (c) and (d) correspond to $h = 40$ MHz and $\delta h \neq 0$.  We use barrier control in (a) and (c), and tilt control in (b) and (d).  We indicate the fitted $T_2$ values in the figures, and provide other parameters in the Supplemental Material \cite{suppl}.
	}
\label{alphaexp}
\end{figure}

Fig.~\ref{alphaexp} shows the results of randomized benchmarking for optimized pulse sequences subject to $1/f^{\alpha_h}$ nuclear noise and $1/f^{\alpha_J}$ charge noise with $\alpha_h = 2.6$ and $\alpha_J=0.7$, as measured in experiments \cite{Medford.12, Rudner.11, Dial.13}. We see that the coherence times $T_2$ are consistently extended with barrier control [panels (a),(c)], and this effect is slightly more pronounced for the case without spin noise, in which $T_2$ has been extended by a factor of 70 [compare panels (a),(b)].  We have also conducted similar calculations with $\alpha_h=\alpha_J=
\alpha$ for $0\le\alpha\le3$, and have presented the results in the Supplemental Material \cite{suppl}.

\begin{figure}
\includegraphics[width=0.9\columnwidth]{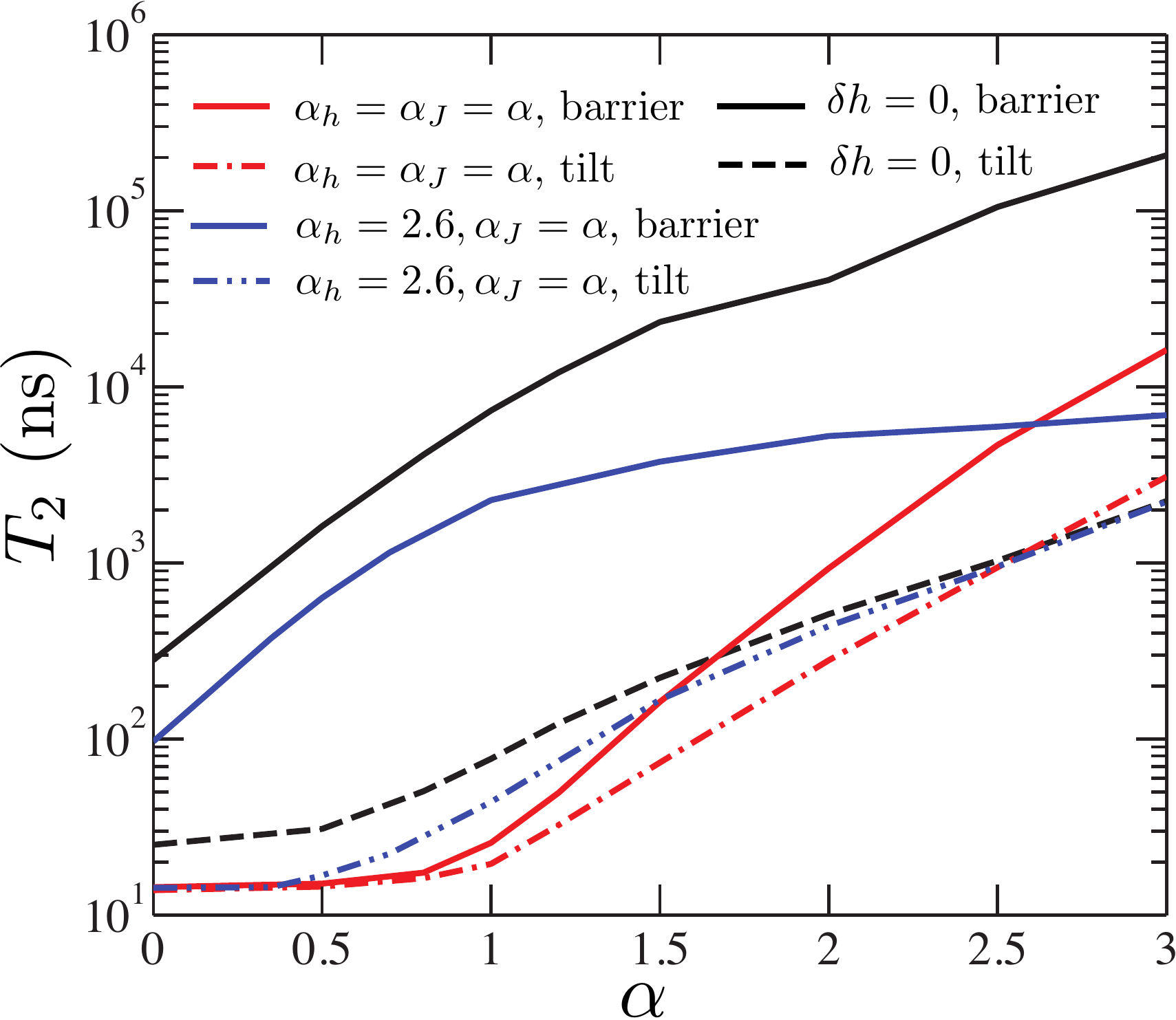}
\caption{$T_{2}$ vs.~$\alpha$ for optimized pulse sequences. Solid lines show the results for barrier control, dashed lines for tilt control. Black lines: nuclear spin noise not present and $\alpha_J=\alpha$ for charge noise. Red lines: $\alpha_h=\alpha_J=\alpha$. Blue lines: $\alpha_h=2.6$ and $\alpha_J=\alpha$.}
\label{ratio}
\end{figure}

Fig.~\ref{ratio} shows the $T_2$ values extracted from randomized benchmarking for optimized pulse sequences under $1/f^\alpha$ noise for $0\le\alpha\le3$. This is the core result of this paper. It is clear from the figure that as $\alpha$ increases from 0 to 3, the $T_2$ times consistently improve for all cases. For the case in which both charge and nuclear spin noise are present and $\alpha_h=\alpha_J$, barrier control gradually outperforms tilt control, with $T_2$ extended by a factor of 10 for $\alpha=3$. In the absence of nuclear spin noise, the $T_2$ values for barrier control are close to 100 times larger than for tilt control for a wide range of $\alpha$. It is remarkable that for the experimentally measured charge noise spectrum with exponent $\alpha_J\approx0.7$ and nuclear spin noise with $\alpha_h=2.6$, the improvement under barrier control is a factor of 70, close to the case without nuclear spin noise. 
Our results show that barrier control combined with our optimized pulses outperforms tilt control for any type of $1/f$ noise and even in the presence of significant nuclear spin noise.

In conclusion, we showed through randomized benchmarking simulations that in the absence of spin noise, symmetric barrier control of singlet-triplet qubits outperforms the traditional tilt control by approximately two orders of magnitude. This result is directly relevant to Si systems, where nuclear spin noise is minimal. However, there is no significant improvement when nuclear spin noise is pronounced and when standard pulse sequences are used to implement gates. To take advantage of barrier control in the presence of nuclear spin noise, we introduced a new family of pulse sequences that implement any single-qubit gate without ever having to tune the exchange interaction down to zero. This reduces noise by speeding up the gates. Thus, barrier control improves coherence times regardless of the level of nuclear spin noise, provided our optimized sequences are used.  In the presence of nuclear spin noise, our optimized pulse sequences outperform standard sequences.

C.Z., X.-C.Y. and X.W. are supported by the Research Grants Council of the Hong Kong Special Administrative Region, China (No. CityU 21300116) and the National Natural Science Foundation of China (No. 11604277).  R.E.T. and S.D.S. are supported by LPS-MPO-CMTC.

%

%%%%%%%%%%%%%%%%%%%%%%%%%%%%%%%%%%
% The main text ends here (note that you need to remove \end{document})
%%%%%%%%%%%%%%%%%%%%%%%%%%%%%%%%%%

%\onecolumngrid
%\newpage

%\begin{widetext}
%\section{Supplementary material}
\vspace{1cm}
\begin{center}
{\bf\large Supplementary material}
\end{center}
\vspace{0.5cm}

\setcounter{secnumdepth}{3}  %enforce numbering of section in PRL template, we keep PRL template because we like the way it labels references
\setcounter{equation}{0}%reset counter
\setcounter{figure}{0}
\setcounter{table}{0}
\renewcommand{\theequation}{S-\arabic{equation}}
\renewcommand{\thefigure}{S\arabic{figure}}
\renewcommand{\thetable}{S-\Roman{table}}
\renewcommand\figurename{Supplementary Figure}
\renewcommand\tablename{Supplementary Table}

\newcolumntype{M}[1]{>{\centering\arraybackslash}m{#1}}
\newcolumntype{N}{@{}m{0pt}@{}}

\makeatletter \renewcommand\@biblabel[1]{[S#1]} \makeatother

%%%%%%%%%%%%%%%%%%%%%%%%%%%%%%%%%%
% The supplementary text starts here
%%%%%%%%%%%%%%%%%%%%%%%%%%%%%%%%%%

In this Supplemental Material we show our randomized benchmarking results for the optimized pulse sequences under $1/f^{\alpha}$ noise for different exponents $\alpha$ that have not been covered in the main text. We also provide tables listing all fitting parameters and relevant noise amplitudes for the numerical simulations that we conducted in this work. In the last section, we give the pulse profiles of all single-qubit Clifford gates used in our randomized benchmarking simulation.

\section{Results of Randomized Benchmarking of optimized sequences under $1/f^{\alpha}$ noise}

In this set of supplementary figures, we show our randomized benchmarking results for the optimized pulse sequences under $1/f^{\alpha}$ noise for noise exponents $\alpha=0$, $0.5$, $1$, $1.5$, $2$, $2.5$, and $3$. Note that in all cases shown in this section, the nuclear noise and charge noise are assumed to share the same exponent, i.e. $\alpha_h=\alpha_J=\alpha$.

\begin{figure}[H]
\centering
	\includegraphics[width=\columnwidth]{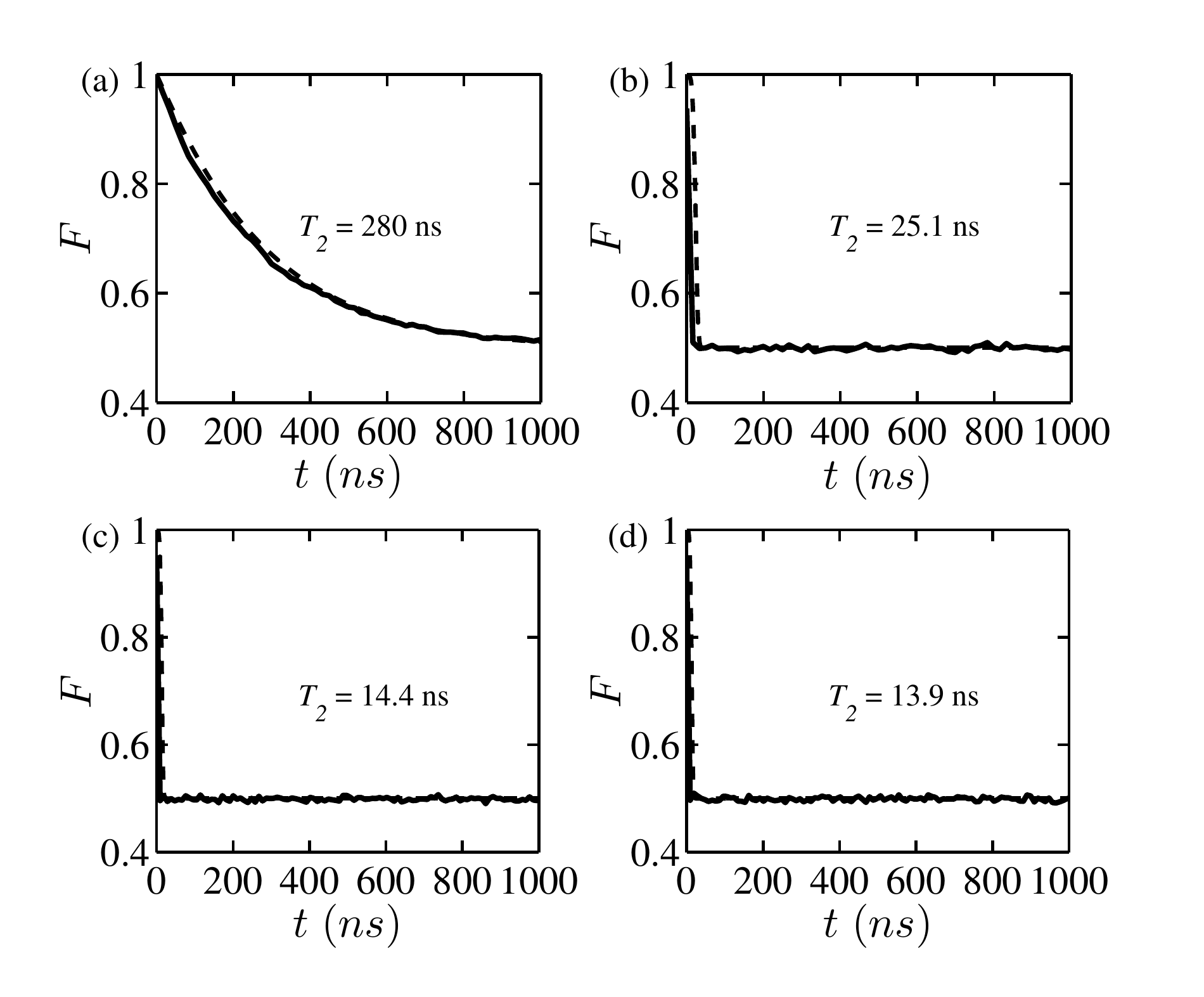}
	\caption{Results of randomized benchmarking with $1/f^0$ noise  (i.e., for  $\alpha_h=\alpha_J = 0$)  for optimized pulses. The solid curves are our numerical results and the dashed curves are fits.  (a) and (b) correspond to $h = 23$ MHz and $\delta h=0$, while (c) and (d) correspond to $h = 40$ MHz and $\delta h \neq 0$.  We use barrier control in (a) and (c), and tilt control in (b) and (d).  We indicate the fitted $T_2$ values in the figures, and provide the $\gamma$ values and the noise amplitudes $A_h$ and $A_J$ in Table~\ref{tab:RBfitting}.}
\label{alpha0}
\end{figure}

\begin{figure}[H]
\centering
	\includegraphics[width=\columnwidth]{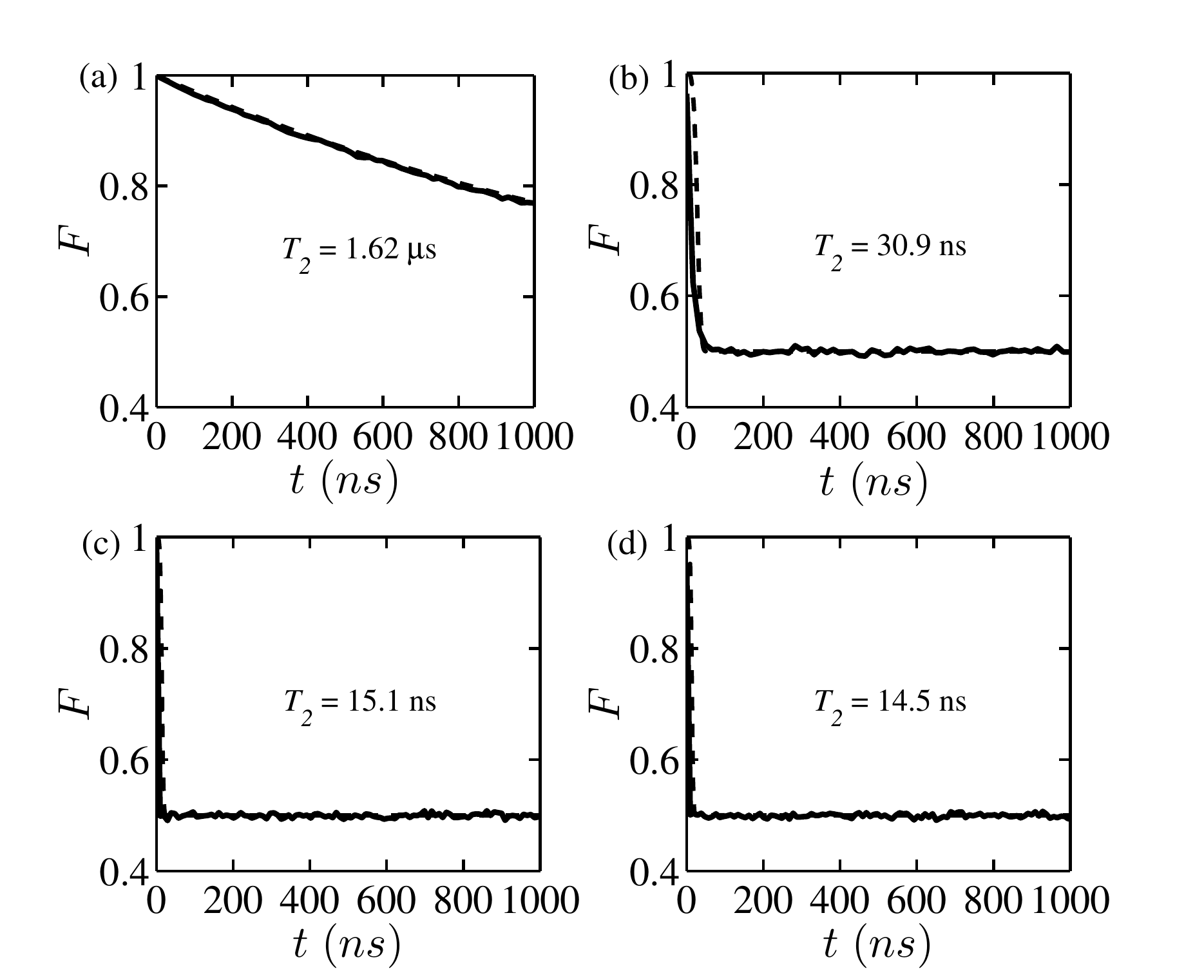}
	\caption{Results of randomized benchmarking with $1/f^{0.5}$ noise  (i.e., for  $\alpha_h=\alpha_J = 0.5$)  for optimized pulses. The solid curves are our numerical results and the dashed curves are fits.  (a) and (b) correspond to $h = 23$ MHz and $\delta h=0$, while (c) and (d) correspond to $h = 40$ MHz and $\delta h \neq 0$.  We use barrier control in (a) and (c), and tilt control in (b) and (d).  We indicate the fitted $T_2$ values in the figures, and provide the $\gamma$ values and the noise amplitudes $A_h$ and $A_J$ in Table~\ref{tab:RBfitting}.}
\label{alpha0p5}
\end{figure}

\begin{figure}[H]
\centering
	\includegraphics[width=0.92\columnwidth]{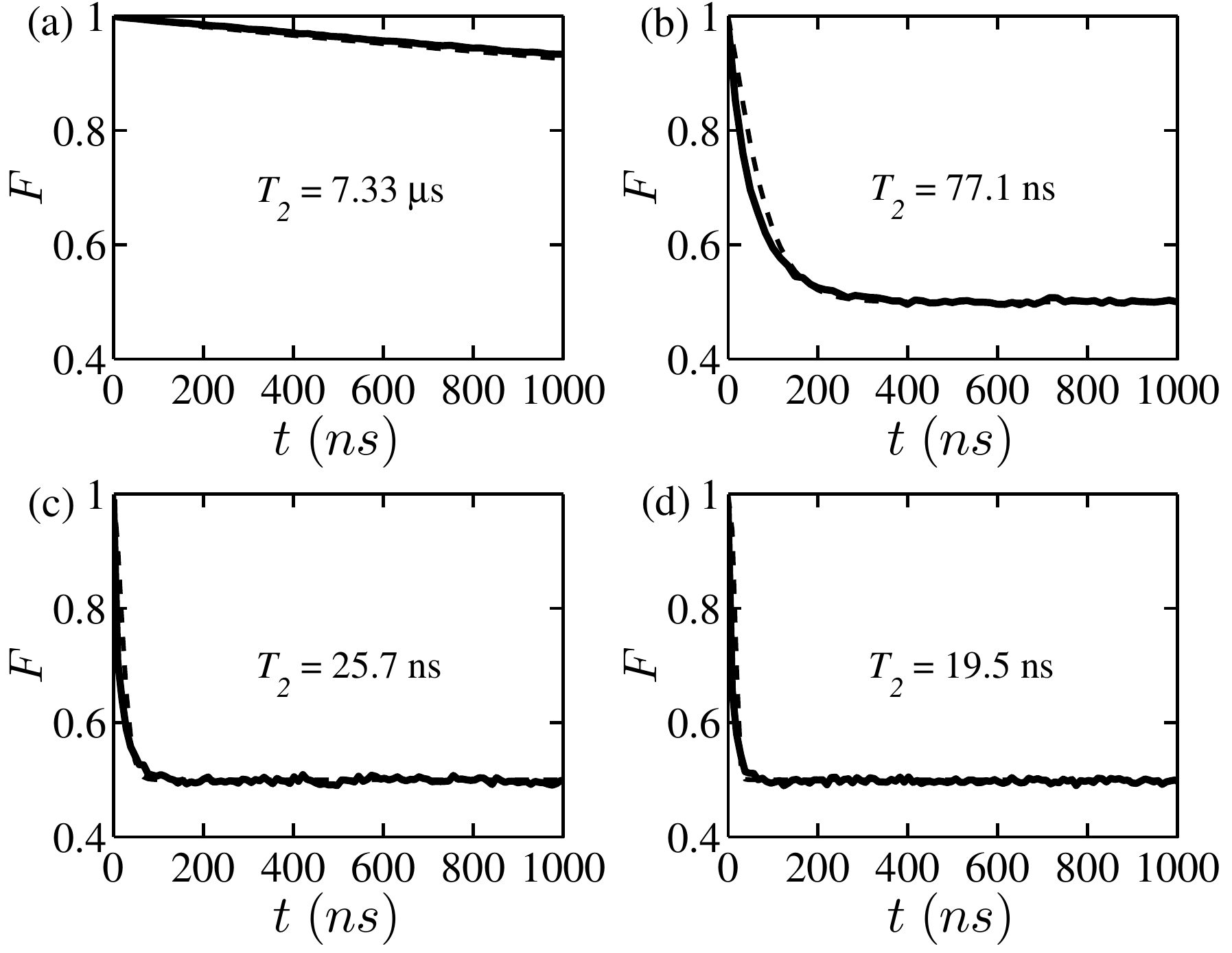}
	\caption{Results of randomized benchmarking with $1/f^{1}$ noise  (i.e., for  $\alpha_h=\alpha_J = 1$)  for optimized pulses. The solid curves are our numerical results and the dashed curves are fits.  (a) and (b) correspond to $h = 23$ MHz and $\delta h=0$, while (c) and (d) correspond to $h = 40$ MHz and $\delta h \neq 0$.  We use barrier control in (a) and (c), and tilt control in (b) and (d).  We indicate the fitted $T_2$ values in the figures, and provide  the $\gamma$ values and the noise amplitudes $A_h$ and $A_J$ in Table~\ref{tab:RBfitting}.}
\label{alpha1}
\end{figure}

\begin{figure}[H]
\centering
	\includegraphics[width=0.92\columnwidth]{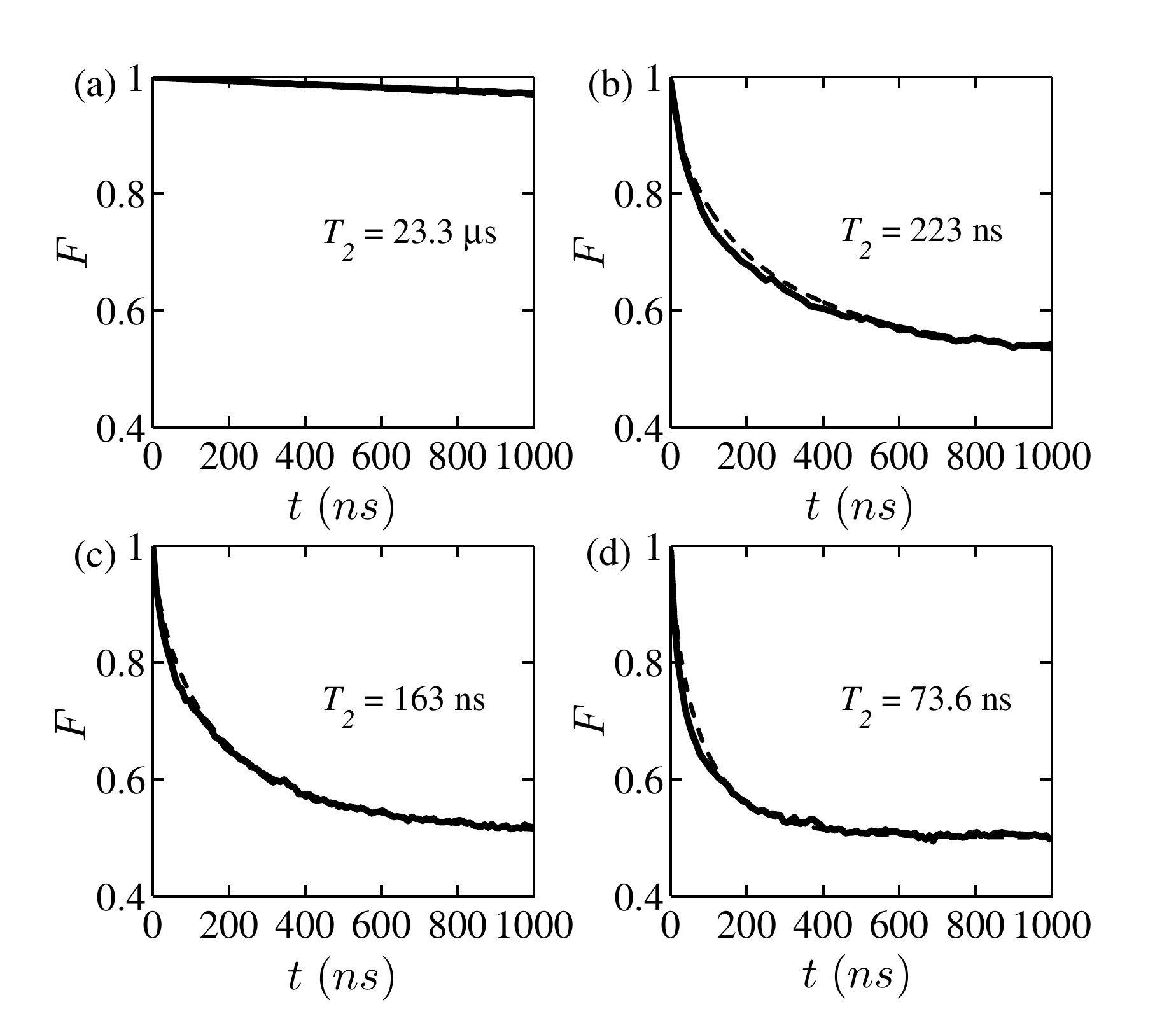}
	\caption{Results of randomized benchmarking with $1/f^{1.5}$ noise  (i.e., for  $\alpha_h=\alpha_J = 1.5$)  for optimized pulses. The solid curves are our numerical results and the dashed curves are fits.  (a) and (b) correspond to $h = 23$ MHz and $\delta h=0$, while (c) and (d) correspond to $h = 40$ MHz and $\delta h \neq 0$.  We use barrier control in (a) and (c), and tilt control in (b) and (d).  We indicate the fitted $T_2$ values in the figures, and provide  the $\gamma$ values and the noise amplitudes $A_h$ and $A_J$ in Table~\ref{tab:RBfitting}.}
\label{alpha1p5}
\end{figure}

\begin{figure}[H]
\centering
	\includegraphics[width=\columnwidth]{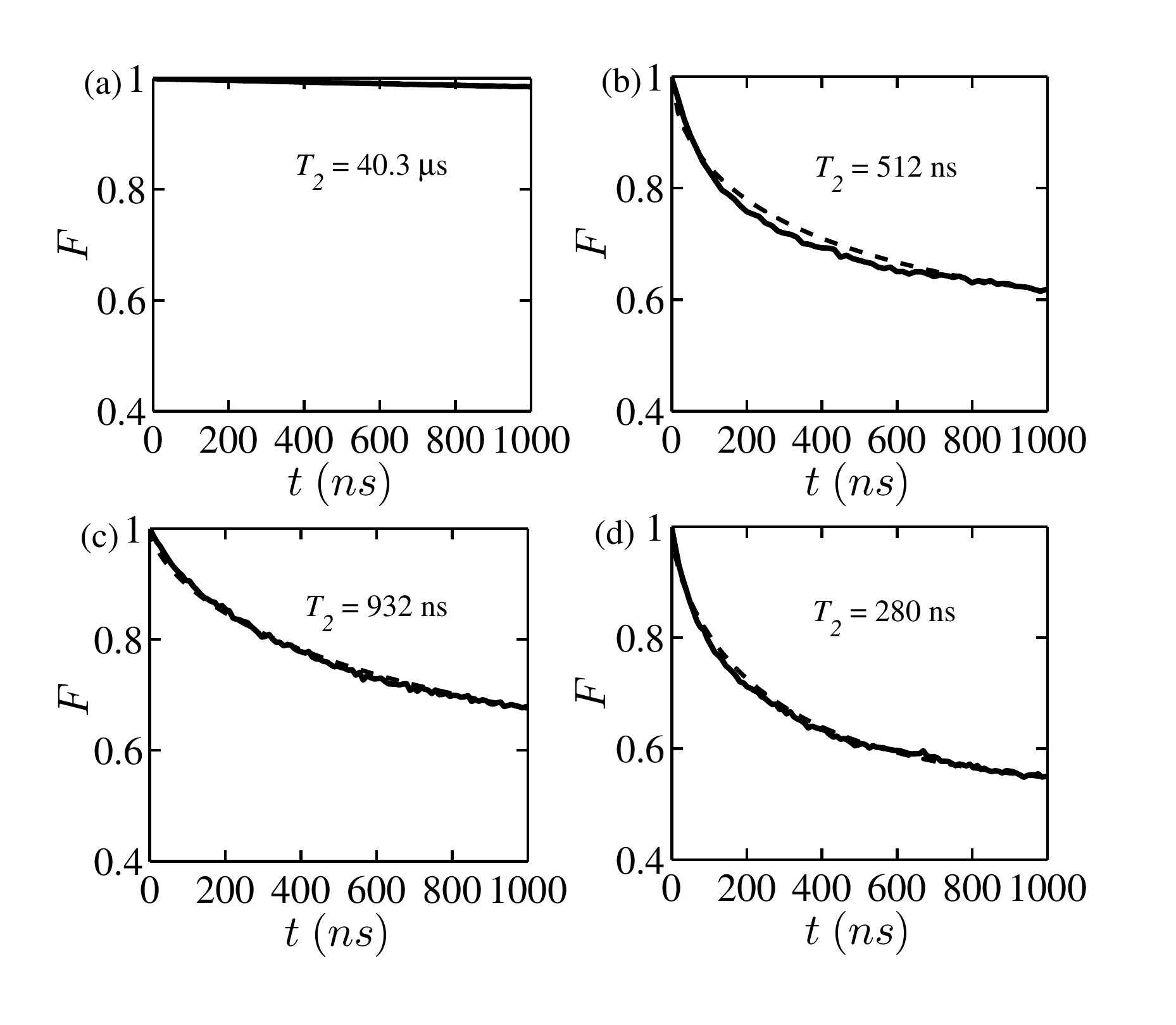}
	\caption{Results of randomized benchmarking with $1/f^2$ noise  (i.e., for  $\alpha_h=\alpha_J = 2$)  for optimized pulses. The solid curves are our numerical results and the dashed curves are fits.  (a) and (b) correspond to $h = 23$ MHz and $\delta h=0$, while (c) and (d) correspond to $h = 40$ MHz and $\delta h \neq 0$.  We use barrier control in (a) and (c), and tilt control in (b) and (d).  We indicate the fitted $T_2$ values in the figures, and provide  the $\gamma$ values and the noise amplitudes $A_h$ and $A_J$ in Table~\ref{tab:RBfitting}.}
\label{alpha2}
\end{figure}

\begin{figure}[H]
\centering
	\includegraphics[width=\columnwidth]{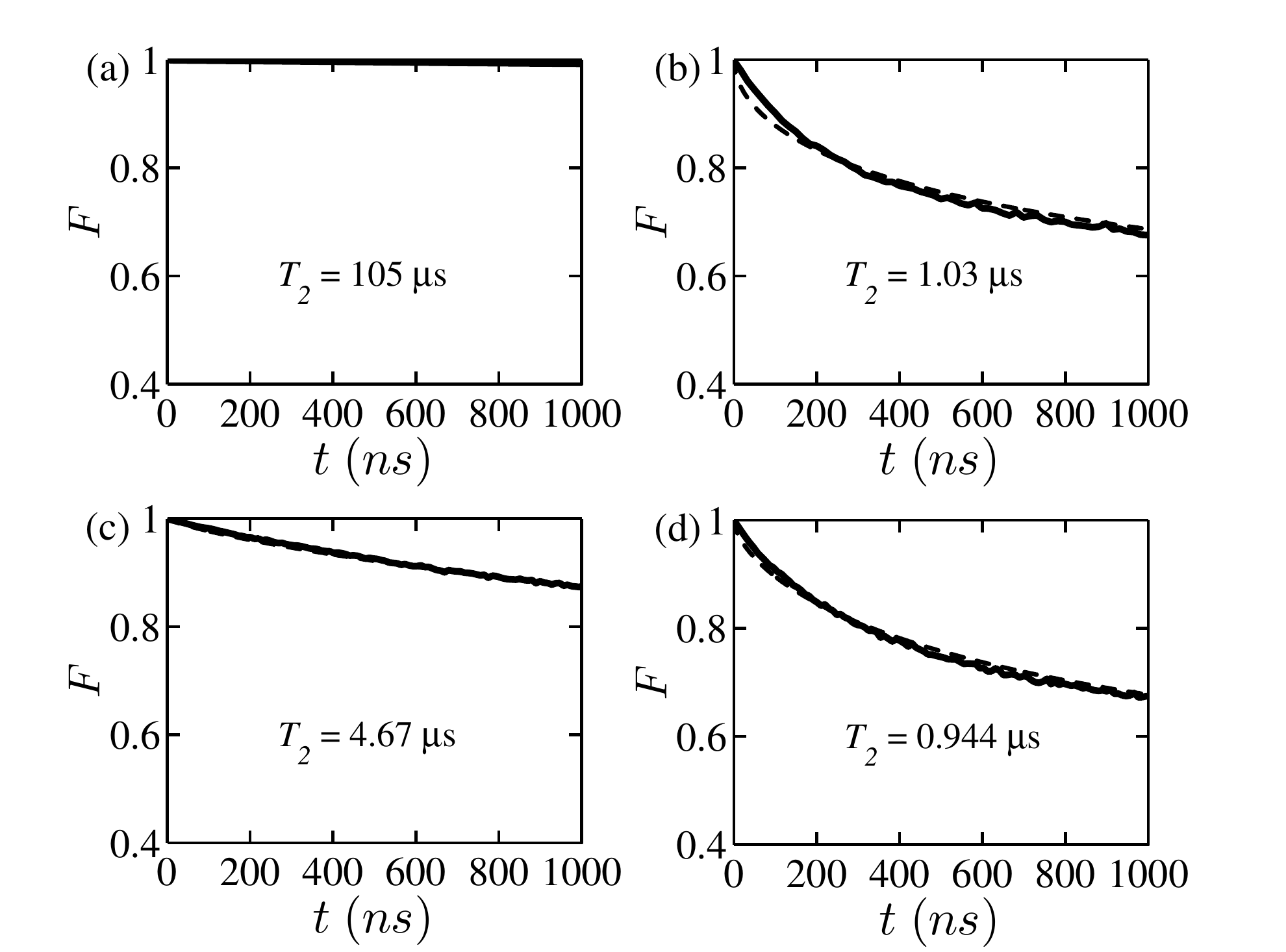}
	\caption{Results of randomized benchmarking with $1/f^{2.5}$ noise  (i.e., for  $\alpha_h=\alpha_J = 2.5$)  for optimized pulses. The solid curves are our numerical results and the dashed curves are fits.  (a) and (b) correspond to $h = 23$ MHz and $\delta h=0$, while (c) and (d) correspond to $h = 40$ MHz and $\delta h \neq 0$.  We use barrier control in (a) and (c), and tilt control in (b) and (d).  We indicate the fitted $T_2$ values in the figures, and provide  the $\gamma$ values and the noise amplitudes $A_h$ and $A_J$ in Table~\ref{tab:RBfitting}.}
\label{alpha2}
\end{figure}

\begin{figure}[H]
\centering
	\includegraphics[width=\columnwidth]{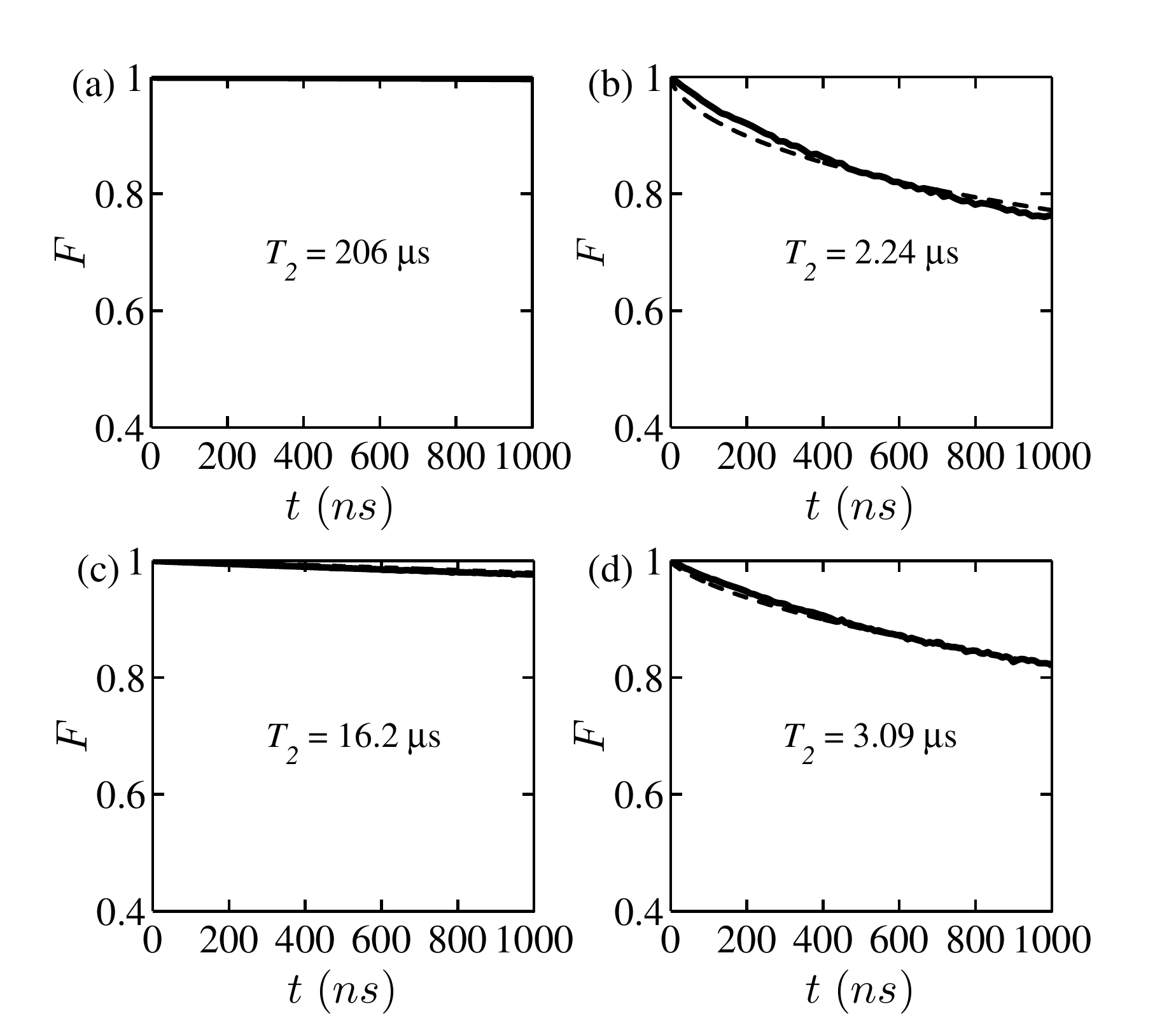}
	\caption{Results of randomized benchmarking with $1/f^3$ noise  (i.e., for  $\alpha_h=\alpha_J = 3$)  for optimized pulses. The solid curves are our numerical results and the dashed curves are fits.  (a) and (b) correspond to $h = 23$ MHz and $\delta h=0$, while (c) and (d) correspond to $h = 40$ MHz and $\delta h \neq 0$.  We use barrier control in (a) and (c), and tilt control in (b) and (d).  We indicate the fitted $T_2$ values in the figures, and provide  the $\gamma$ values and the noise amplitudes $A_h$ and $A_J$ in Table~\ref{tab:RBfitting}.}
\label{alpha3}
\end{figure}

%\onecolumngrid
%\pagebreak

\section{Fitting Parameters and noise amplitudes of all results presented}

In the tables on the next page, we give all fitting parameters and noise amplitudes that have been extracted or used in our numerical randomized benchmarking simulations. In most cases, fitting to $\left\{1+\exp\left[-(t/T_2)^\gamma\right]\right\}/2$ is sufficient. In this case a single $\gamma$ value is provided in the corresponding entries of the table. However in some cases a better fit is achieved using a summation of two exponentials $\left\{2+\exp\left[-(t/T_2)^{\gamma_1}\right]+\exp\left[-(t/T_2)^{\gamma_2}\right]\right\}/4$, in which case both $\gamma_1$ and $\gamma_2$ are given.

\begin{table*}[!htbp]
\footnotesize		
\centering
	\begin{tabular}{|M{115pt}|M{80pt}|M{135pt}|M{135pt}|N}
		\hline
		Nuclear noise & Pulse sequence & Barrier control & Tilt control &\\[8pt]	
		\hline
		\hline
		\multirow{2}{*}[-0.5em]{$h = 23$ Mhz, $\sigma_{h} =0$} & Optimized  & $T_{2} = 17.6  {\rm \mu s}$, $\gamma = 0.702$  &  $T_{2} =0.160  {\rm \mu s},$ $\gamma = 0.310$  &\\[10pt]
		\cline{2-4}  
		& Unoptimized &  $T_{2} = 733  {\rm \mu s},$ $\gamma = 0.843$  &  $T_{2} = 6.81 {\rm \mu s},$ $\gamma = 0.532$ &\\[8pt]
		\cline{2-4}
		\hline 
		\multirow{2}{*}[-0.5em]{$h = 40$ MHz, $\sigma_{h}  = 11.5$ MHz} & Optimized  & $T_{2} = 183  {\rm ns}$, $\gamma_1 = 0.265$,  $\gamma_2 = 0.561$ &  $T_{2} =50.0  {\rm ns},$ $\gamma_1 = 1.37$, $\gamma_2 = 0.345$  &\\[8pt]
		\cline{2-4}  
		& Unoptimized &  $T_{2} = 52.2  {\rm ns},$ $\gamma_1 = 0.208$, $\gamma_2 = 0.909$  &  $T_{2} = 45.1 {\rm ns},$ $\gamma_1 = 0.300$, $\gamma_2 = 1.00$ &\\[8pt]
		\cline{2-4}
		\hline 
		\multirow{2}{*}[-0.5em]{$h = 40$ MHz, $\sigma_{h}  = 23$ MHz} & Optimized  & $T_{2} = 58.4  {\rm \mu s}$, $\gamma_1 = 1.04$, $\gamma_2 = 0.256$  &  $T_{2} =27.8  {\rm ns},$ $\gamma_1 = 0.362$, $\gamma_2 = 2.56$  &\\[8pt]
		\cline{2-4}  
		& Unoptimized &  $T_{2} = 22.3  {\rm \mu s},$ $\gamma_1 = 1.00$,$\gamma_2 = 0.258$  &  $T_{2} = 22.2 {\rm ns},$ $\gamma_1 = 1.00$, $\gamma_2 = 0.337$ &\\[8pt]
		\cline{2-4}
		\hline 				
	\end{tabular}
	\caption{Fitting results for the randomized benchmarking data presented in Fig.~\ref{fig:static} of the main text.}\label{tab:three}
\end{table*}

\begin{table*}[!htbp]
\footnotesize	
\centering
\begin{tabular}{|M{40pt}|M{210pt}|M{210pt}|N}
\hline
 $\delta h$ & Barrier control & Tilt control &\\[8pt]		
\hline
\hline
\multirow{2}{*}[0.5em]{$\delta h = 0$} & $T_{2} = 3.00 {\rm \mu s}$, $\gamma = 1.01,$ $\lg A_{J} = -4.13,$ $A_{h} = 0$ &  $T_{2} = 42.0 {\rm ns},$ $\gamma = 1.98,$ $\lg A_{J} = -1.89,$ $A_h = 0$ &\\[8pt]
\cline{1-3}  
\multirow{2}{*}[0.5em]{$\delta h \neq0$} &  $T_{2} = 1.15 {\rm \mu s},$ $\gamma = 1.03,$ $\lg A_{J} = -4.06,$ $\lg A_h = -5.53$ &  $T_{2} = 22.3 {\rm ns},$ $\gamma = 2.14,$ $\lg A_{J} = -1.82,$ $\lg A_h = -5.53$ &\\[8pt]
\cline{1-3}
\hline 
\end{tabular}
\caption{Fitting parameters and noise amplitudes for randomized benchmarking data presented in Fig.~\ref{alphaexp} of the main text.}\label{tab:three}
\end{table*}

\begin{table*}[!htbp]
\footnotesize	
\centering
 \begin{tabular}{|M{16pt}|M{30pt}|M{220pt}|M{220pt}|N}
\hline
 $\alpha$ & $\delta h$ & Barrier control & Tilt control &\\[8pt]

\hline
\hline
\multirow{2}{*}[-0.5em]{$0$} & $\delta h = 0$  & $T_{2} = 280  {\rm ns}$, $\gamma_1 = 1.05,$ $\gamma_2 = 1.05,$ $\lg A_{J} = -2.86,$ $A_h = 0$ &  $T_{2} =25.1  {\rm ns},$ $\gamma_1 = 5.96,$ $\gamma_2 = 3.99,$ $\lg A_{J} = -0.618,$ $A_h = 0$ &\\[8pt]
 \cline{2-4}  
 & $\delta h \neq0$ &  $T_{2} = 14.4 {\rm ns},$ $\gamma_1 = 6.00,$ $\gamma_2 = 4.00,$ $\lg A_{J} = -2.62,$ $\lg A_h = 2.66$ &  $T_{2} = 13.9 {\rm ns},$ $\gamma_1 = 6.00,$ $\gamma_2 = 4.00,$  $\lg A_{J} = -0.377,$ $\lg A_h = 2.66$ &\\[8pt]
\cline{2-4}
\hline 

\multirow{2}{*}[-0.5em]{$0.5$} & $\delta h = 0$  & $T_{2} = 1.62  {\rm \mu s}$, $\gamma = 1.01,$ $\lg A_{J} = -3.75,$ $A_h = 0$ &  $T_{2} = 30.9 {\rm ns},$ $\gamma_1 = 6.00,$ $\gamma_2 =2.61,$ $\lg A_{J} = -1.51,$ $A_h = 0$ &\\[8pt]
 \cline{2-4}  
 & $\delta h \neq0$ &  $T_{2} = 15.1 {\rm ns},$ $\gamma_1 = 6.00,$ $\gamma_2 = 4.00,$ $\lg A_{J} = -3.63,$ $\lg A_h = 1.18$ &  $T_{2} = 14.5 {\rm ns},$ $\gamma_1 = 6.00,$  $\gamma_2 = 4.00,$ $\lg A_{J} = -1.39,$ $\lg A_h = 1.18$ &\\[8pt]
\cline{2-4}
\hline 

\multirow{2}{*}[-0.5em]{$1$} & $\delta h = 0$  & $T_{2} = 7.33 {\rm \mu s}$, $\gamma = 0.915,$ $\lg A_{J} = -4.72,$ $A_h = 0$ &  $T_{2} = 77.1 {\rm ns},$ $\gamma = 1.21,$ $\lg A_{J} = -2.48,$ $A_h = 0$ &\\[8pt]
  \cline{2-4}  
  & $\delta h \neq0$ &  $T_{2} = 25.7 {\rm ns},$ $\gamma = 1.61,$ $\lg A_{J} = -4.72,$ $\lg A_h = -0.346$ &  $T_{2} = 19.5 {\rm ns},$ $\gamma = 2.76,$ $\lg A_{J} = -2.48,$ $\lg A_h = -0.346$ &\\[8pt]
\cline{2-4}
\hline

\multirow{2}{*}[-0.5em]{$1.5$} & $\delta h = 0$  & $T_{2} = 23.3 {\rm \mu s}$, $\gamma = 0.862,$ $\lg A_{J} = -5.77,$ $A_h = 0$ &  $T_{2} = 223 {\rm ns},$ $\gamma = 0.659,$ $\lg A_{J} = -3.52,$ $A_h = 0$ &\\[8pt]
  \cline{2-4}  
  & $\delta h \neq0$ &  $T_{2} = 163 {\rm ns},$ $\gamma = 0.700,$ $\lg A_{J} = -5.89,$ $\lg A_h = -1.92$ &  $T_{2} = 73.6 {\rm ns},$ $\gamma = 0.743,$ $\lg A_{J} = -3.64,$ $\lg A_h = -1.92$ &\\[8pt]
\cline{2-4}
\hline  

 \multirow{2}{*}[-0.5em]{$2$} & $\delta h = 0$  & $T_{2} = 40.3 {\rm \mu s}$, $\gamma = 0.931,$ $\lg A_{J} = -6.88,$ $A_h = 0$ &  $T_{2} = 512 {\rm ns},$ $\gamma_1 = 0.576,$ $\gamma_2 = 0.576,$ $\lg A_{J} = -4.64,$ $A_h = 0$ &\\[8pt]
   \cline{2-4}  
   & $\delta h \neq0$ &  $T_{2} = 932 {\rm ns},$ $\gamma = 0.657,$ $\lg A_{J} = -7.12,$ $\lg A_h = -3.54$ &  $T_{2} = 280 {\rm ns},$ $\gamma_1 = 0.532,$ $\gamma_2 = 0.851,$ $\lg A_{J} = -4.88,$ $\lg A_h = -3.54$ &\\[8pt]
\cline{2-4}
\hline  

\multirow{2}{*}[-0.5em]{$2.5$} & $\delta h = 0$  & $T_{2} = 105 {\rm \mu s}$, $\gamma = 0.895,$ $\lg A_{J} = -8.06,$ $A_h = 0$ &  $T_{2} = 1.03 {\rm \mu s},$ $\gamma = 0.549,$ $\lg A_{J} = -5.82,$ $A_h = 0$ &\\[8pt]
   \cline{2-4}  
   & $\delta h \neq0$ &  $T_{2} = 4.67 {\rm \mu s},$ $\gamma = 0.792,$ $\lg A_{J} = -8.42,$ $\lg A_h = -5.20$ &  $T_{2} = 0.944 {\rm \mu s},$ $\gamma = 0.647,$ $\lg A_{J} = -6.18,$ $\lg A_h = -5.20$ &\\[8pt]
\cline{2-4}
\hline 

\multirow{2}{*}[-0.5em]{$3$} & $\delta h = 0$  & $T_{2} = 206 {\rm \mu s}$, $\gamma = 0.979,$ $\lg A_{J} = -9.27,$ $A_h = 0$ &  $T_{2} = 2.24 {\rm \mu s},$ $\gamma = 0.615,$ $\lg A_{J} = -7.03,$ $A_h = 0$ &\\[8pt]
 \cline{2-4}  
 & $\delta h \neq0$ &  $T_{2} = 16.2 {\rm \mu s},$ $\gamma = 1.14,$ $\lg A_{J} = -9.75,$ $\lg A_h = -6.88$ &  $T_{2} = 3.09 {\rm \mu s},$ $\gamma = 0.730,$ $\lg A_{J} = -7.51,$ $\lg A_h = -6.88$ &\\[8pt]
\cline{2-4}
\hline 
 
 \end{tabular}
 \caption{Fitting parameters and noise amplitudes for randomized benchmarking data of our optimized pulse sequences presented in Supplementary Figures~\ref{alpha0} through \ref{alpha3} in this document.}\label{tab:RBfitting}
\end{table*}

\onecolumngrid
%\pagebreak
\section{Single-qubit Clifford gates used in the randomized benchmarking}

In Supplementary Fig.~\ref{Cliffords} we present the pulse profiles of all single-qubit Clifford gates used in our simulation of  randomized benchmarking. For the identity operation, the optimized and unoptimized pulses are the same; for $x$-rotations, the optimized pulses are longer because they do not allow $J=0$ while keeping $J$ at zero is the most direct way to achieve an $x$-rotation.  The pulse lengths of  $z$-rotations are substantially reduced for optimized ones as compared to the unoptimized ones, while for all other rotations, the durations of the optimized pulses are reduced by about 40\% to 60\%.

\begin{figure}[H]
\centering
	\includegraphics[width=\columnwidth]{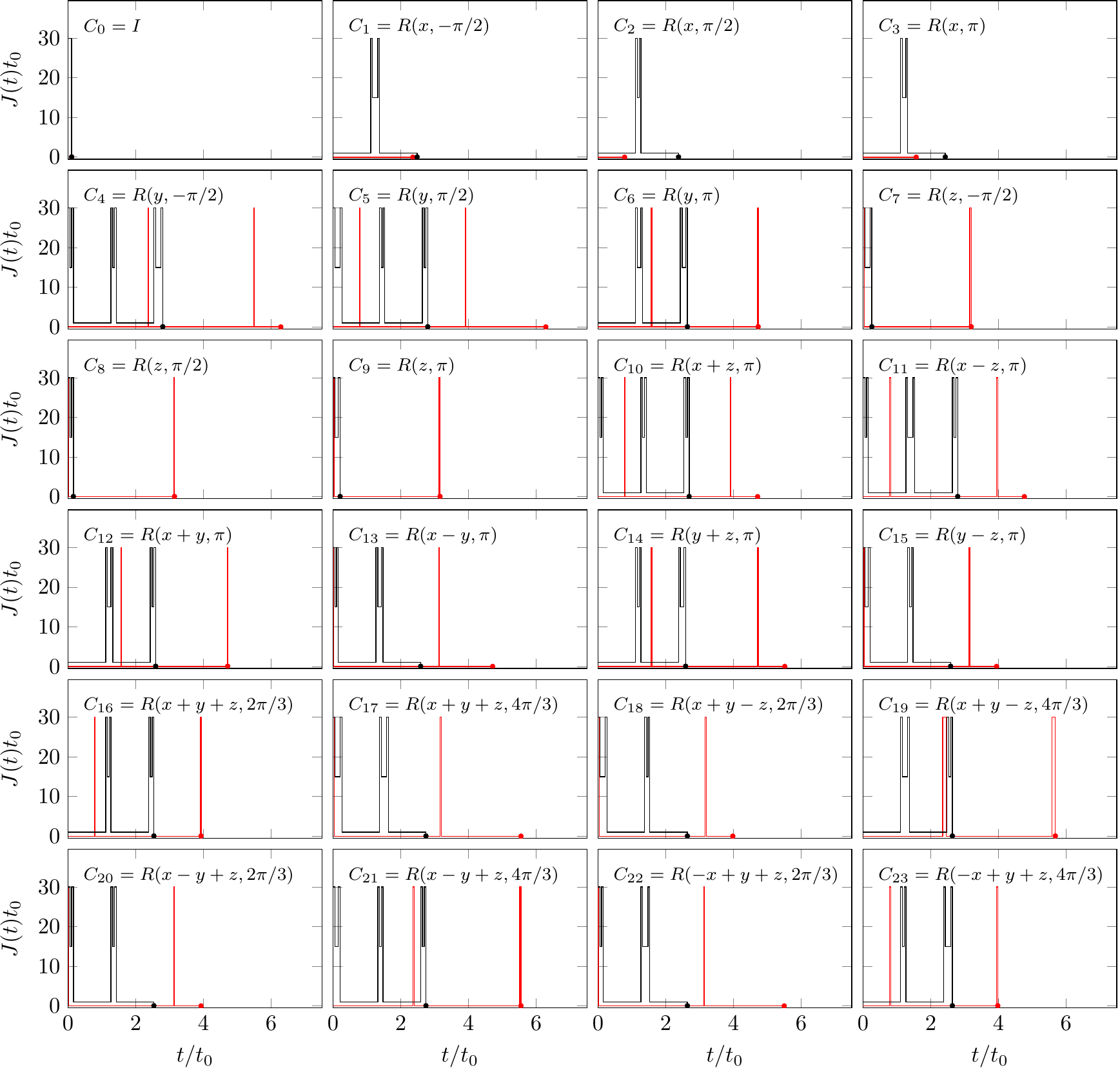}
	\caption{Pulse profiles of all single-qubit Clifford gates used in the randomized benchmarking simulation. Red lines: unoptimized pulses; black lines: optimized ones.}
\label{Cliffords}
\end{figure}


\begin{thebibliography}{47}
\bibitem [{xro()}]{xrot}%
  \BibitemOpen
  \href@noop {} {}\bibinfo {note} {For $x$ rotations, although it is desirable
  to use Eq.~\eqref{eq:Robert} with $x$ and $z$ interchanged, the implied $J$
  value will be less than $h$ making the pulses inefficient.}\BibitemShut
  {Stop}%
\end{thebibliography}
\end{document}